\pdfoutput=1
\documentclass[a4paper,fleqn,usenatbib]{aa}

\usepackage{graphicx}
\usepackage{txfonts}
\usepackage{natbib}
\usepackage[]{hyperref}

\hypersetup{
    bookmarks=true,                                   
    pdftitle={}, 
    pdfauthor={}, 
    colorlinks=true,                                  
    linkcolor=blue,                                   
    citecolor=blue,                                   
    urlcolor=blue                                     
}

\newcommand{\GD}{$\gamma$ Dor\,}
\newcommand{\Kepler}{{\it Kepler}\,}

\begin{document}

\title{$\gamma$ Doradus stars as test of angular momentum transport models} 
\subtitle{}

\author{R-M. Ouazzani\inst{1} 
  \and 
  J.P. Marques\inst{2}
  \and 
  M-J. Goupil\inst{1} 
  \and 
  S. Christophe\inst{1}
  \and 
  V. Antoci\inst{3}
  \and 
  S.J.A.J. Salmon\inst{4}
}
\institute{LESIA, Observatoire de Paris, PSL Research University, CNRS, Sorbonne Universit\'es, UPMC Univ. Paris 06, Univ. Paris Diderot,
  Sorbonne Paris Cit\'e, 5 place Jules Janssen, 92195 Meudon, France
  \and
  Univ. Paris-Sud, Institut d'Astrophysique Spatiale, UMR 8617, CNRS, B\^atiment 121, 91405, Orsay Cedex, France
  \and
  Stellar Astrophysics Centre, Department of Physics and Astronomy, Aarhus University, Ny Munkegade 120, DK-8000 Aarhus C, Denmark
  \and
  STAR Institute, Universit\'e de Li\`ege, All\'ee du 6 ao\^ut 19, 4000 Li\`ege, Belgium
}
\date{Draft: \today; Received xxx; accepted xxx}

\abstract{Helioseismology and asteroseismology of red giant stars have shown that distribution of angular momentum in stellar interiors, and its evolution with time remains an open issue in stellar physics. Owing to the unprecedented quality and long baseline of \Kepler photometry, we are able to seismically infer internal rotation rates in $\gamma$ Doradus stars, which provide the main-sequence counterpart to the red-giants puzzle.}
         {Here, we confront these internal rotation rates to stellar evolution models which account for rotationally induced transport of angular momentum, in order to test angular momentum transport mechanisms.}
         {On the one hand, we used a stellar model-independent method developed by Christophe et al. in order to obtain accurate, seismically inferred, buoyancy radii and near-core rotation for 37 $\gamma$ Doradus stars observed by \Kepler. We show that the stellar buoyancy radius can be used as a reliable evolution indicator for field stars on the main sequence. On the other hand, we computed rotating evolutionary models of intermediate-mass stars including internal transport of angular momentum in radiative zones, following the formalism developed by Zahn and Maeder, with the CESTAM code. This code calculates the rotational history of stars from the birth line to the tip of the RGB. The initial angular momentum content has to be set initially, which is done here by fitting rotation periods in young stellar clusters.}
         {We show a clear disagreement between the near-core rotation rates measured in the sample and the rotation rates obtained from the evolutionary models including rotationally induced transport of angular momentum following Zahn (1992). These results show a disagreement similar to that of the Sun and red giant stars in the considered mass range. This suggests the existence of missing mechanisms responsible for the braking of the core before and along the main sequence. The efficiency of the missing mechanisms is investigated.}
         {The transport of angular momentum as formalized by Zahn and Maeder cannot explain the measurements of near-core rotation in main-sequence intermediate-mass stars we have at hand.}


\titlerunning{$\gamma$ Doradus stars as test of angular momentum transport}
\authorrunning{Ouazzani et al.}

\maketitle

\section{Introduction}
Distribution of angular momentum in stellar interiors, and its evolution with time remains an open issue in stellar physics. The first reason is that internal angular momentum distribution has been poorly constrained by observations until recently. The tightest constraint on internal angular momentum is provided by seismic measurements of rotation profiles, thanks to helio- and asteroseismology.

One recent success of this approach was obtained for sub-giant and red giant stars. Being in late stages of stellar evolution, these stars have a highly condensed core, and hence present non-radial mixed modes of oscillation. These modes are of particular interest for the determination of the rotation profile throughout the star, as they carry information on the star's innermost layers and are detectable at the surface. The NASA \Kepler spacecraft has allowed a significant leap forward providing exquisite seismic observations of thousands of such stars. Their rotationally split multiplets give access to near-core rotation rates \citep{Beck2012,Deheuvels2012,Mosser2012b}, which were found to be much slower than predicted by the current models including physically motivated angular momentum transport mechanisms \citep{Eggenberger2012,Marques2013,Cantiello2014,Fuller2014,Belkacem2015b}. The disagreement with observations points toward the existence of missing mechanisms which would extract angular momentum from the core. Currently, the community effort is put into two complementary approaches. On the one hand increasing the number of measurements and inferring correlations with other stellar properties \citep{Gehan2016}, and on the other hand developing parametrized models of core-to-envelope coupling processes \citep[in terms of time-scales, efficiency, etc., see for instance][]{Eggenberger2017}.

A crucial path forward is to get an insight on the angular momentum distribution at earlier stages of evolution, i.e. on the main sequence. Unfortunately, solar-like stars on the main sequence only exhibit pressure modes to the level of detection, and these modes probe their superficial layers. The only solar-like star on the main sequence for which seismology provides information on the deeper layers is the Sun. The long-lived satellite SoHO has successfully provided seismic observations of the Sun's interior, which allowed the inversion of the solar internal rotation profile \citep[e.g.][]{Schou1998,Garcia2007,Fossat2017}. As with red giant stars, these seismic measurements first questioned the understanding of angular momentum transport mechanisms.

In order to obtain inner rotation profiles for a number of stars on the main sequence, and for stars that are progenitors of red giants stars, one has to consider slightly more massive stars than the Sun. Among these stars, a quite interesting sample is that if $\gamma$ Doradus ($\gamma$ Dor) stars which are late A- to early F-type stars with masses ranging from 1.3 to 2 M$_{\odot}$. They burn hydrogen in their convective cores, surrounded by a radiative region where a shallow convective layer subsists. Such shallow convective layers give birth to gravity oscillation modes (g modes) excited by the convective blocking mechanism \citep{Dupret2005}. Due to the structure of these stars, the detected oscillations are able to probe the deep internal region of these stars. With periods typically of the order of one day, these oscillations were extremely difficult to detect from ground. One had to wait for the four years of nearly continuous observations from \Kepler to obtain data of sufficient quality in order to perform seismic studies of these stars.
Moreover, unlike solar-type stars, the convective envelopes in \GD stars are too shallow to generate a magnetic field able to act as a magnetic torque and slow down their surfaces \citep{Schatzman1962}. For that reason these stars have projected rotation velocities of around 100 km.s$^{-1}$ in average \citep{Abt1995}. These high-rotation velocities have long hampered the interpretation of \GD stars seismology.

The first \Kepler \GD stars analysed were slow rotators, for which it was still possible to retrieve rotationally split g-modes. Hence, the first internal rotation periods, by choice of method, were found very slow \citep[of the order of a few tens of days, see][]{Kurtz2014,Saio2015,Schmid2015,Keen2015,Murphy2016}. New methods had to be developed before the rapid rotators could be analysed. Based on patterns in the oscillation periods of their seismic spectra, it was finally possible to infer the near-core rotation rates, for rapid rotators \citep{VanReeth2016,Ouazzani2017}. In particular, \cite{Ouazzani2017} established a one-to-one relation between an observable of the periodogram and the inner rotation rate, which is valid for the whole \GD instability strip. These studies allowed to find near-core rotation frequencies for these stars ranging from 5 to 25 $\mu$Hz -periods of 0.5 to 2.5 days- for dozens of these stars.

The present study is a first attempt to confront these findings with the models of angular momentum (AM) evolution in intermediate-mass stars. In particular, we make use of the stellar evolution code {\sc cestam} \citep{Marques2013}, which models AM evolution from the stellar birth-line to the tip of the red-giant branch. These models are presented in Sect. \ref{S_model}, where the choices for angular momentum transport processes are explained. Such calculations require the definition of appropriate initial conditions, which set the initial AM content. These initial conditions are chosen using observations of young stellar forming regions as a constraint. This will constitute the third section (\ref{S_init}). In order to follow AM distribution along evolution, we define a core-averaged property, the buoyancy radius, and show that it can be readily used as an age indicator on the main sequence of these stars (Sect. \ref{S_BR}). Finally, we report on the results in Sect. \ref{S_results}, before discussing them, and drawing conclusions in Sect. \ref{S_cp}.

 \section{Angular momentum evolution modelling}
 \label{S_model}
 \subsection{Angular momentum transport}
 \label{Ss_AM}
 The stellar models used in this study have been computed with the {\sc cestam} evolution code which originates from the {\sc cesam} code \cite{Morel1997,Morel2008}, where rotationally induced transport has been implemented \citep{Marques2013}. In convective zones, although there is differential rotation in latitude, the mean rotation at a given radius weakly depends on the radius. Therefore, in {\sc cestam} models, that are one-dimensional, we assumed that convective zones rotate as solid bodies. In radiative zones, the transport of angular momentum and chemical elements is modelled following the formalism of \cite{Zahn1992} (hereafter Z92), refined in \cite{Maeder1998}. According to these studies, because of the stable stratification in radiative zones, turbulence is much stronger in the horizontal than in the vertical direction. Thus we make the hypothesis of shellular rotation -i.e. the rotation rate is almost constant on isobars.

 The transport of angular momentum obeys an advection-diffusion equation:
 \begin{equation}
   \label{eq:AMT}
   \rho \frac{\rm d}{{\rm d}t}\left( r^2 \Omega\right) = \frac{1}{5 r^2} \frac{\partial }{\partial r} \left(\rho r^4 \Omega U_2  \right) + \frac{1}{r^2} \frac{\partial }{\partial r} \left(\rho r^4 \nu_{\rm v}  \frac{\partial \Omega}{\partial r}\right), 
 \end{equation}
 where $\rho$ is the stellar density, r the radial coordinate, $\Omega$ the rotation angular velocity, $U_2$ is the vertical component of the meridional circulation, and $\nu_{\rm v}$ is the vertical component of the turbulent viscosity.
 Meridional circulation components are calculated following \cite{Maeder1998}. The shear induced turbulence is considered to be a highly anisotropic diffusive process. For the diffusion coefficients, we chose the prescriptions for horizontal turbulent diffusion coefficient from \cite{Mathis2004}, and for the vertical one \cite{Talon1997}. According to \cite{Amard2016}, this combination, together with \cite{Matt2012,Matt2015} for the loss of AM by magnetized winds, give the best fit to rotation periods of solar-like stars in clusters.
 In the current study, we did not consider loss of AM at the surface. The external convective zone in $\gamma$ Dor stars being much shallower than that of solar-like stars, it has been assumed that generation of magnetic field by a dynamo-like process would be inefficient.
 Turbulence in the vertical direction also mixes chemical elements. This mixing is further enhanced by the large-scale meridional circulation coupled to a strong horizontal turbulence \citep{Chaboyer1992}.
 As a result, the equation of the chemical composition evolution follows:
 \begin{equation}
   \label{eq:chem}
   \frac{{\rm d}X_i}{{\rm d}t} = \frac{\partial }{\partial m} \left[ \left( 4 \pi r^2 \rho \right)^2 \left( D_v + D_{eff} \right) \frac{\partial X_i}{\partial m} \right] + \left.\frac{{\rm d} X_i}{{\rm d} t}\right)_{\rm nucl}\,,
 \end{equation}
where $X_i$ is the abundance by mass of the $i$th nuclear species, $D_v = \nu_v$ is the vertical component of turbulent diffusion, and $D_{eff}$ the vertical diffusivity and the diffusion coefficient associated with meridional circulation. Note that in the present study, we neglect atomic diffusion, whose effects should be small compared to turbulent diffusion induced by differential rotation. In the following, for shortness, we will refer to the formalism and prescriptions mentioned in this section as Z92.

\subsection{Stellar models}
These prescriptions have been used in order to compute stellar models for masses between 1.4 and 1.8M$_{\odot}$. At each metallicity, we have derived the helium mass fraction using a primordial Y$_{\rm p} = 0.248$ \citep{Peimbert2007}, and a helium-to-metal enrichment ratio of $\frac{\Delta Y}{\Delta Z} = 2.1 \pm 0.9$ \citep{Casagrande2007}. We adopted the {\sc agss09} solar metal mixture \citep{Asplund2009} and corresponding opacity tables obtained with {\sc opal} opacities \citep{Iglesias1996}, completed at low temperature ($\log T < 4.1$) with \cite{Alexander1994} opacity tables. We used {\sc opal} equation of state \citep{Rogers1996}. We used {\sc nacre} nuclear reaction rates of \cite{Angulo1999} except for the $^{14}N + p$ reaction, for which we used the reaction rates given in \cite{Formicola2004}. The Schwarzschild criterion was used to determine convective instability. Convection was treated using the mixing-length theory (MLT) formalism \citep{Bohm-Vitense1958} with a parameter $\alpha_{\rm MLT}=1.70$. The centrifugal acceleration is taken into account by adding the average centrifugal acceleration $2\Omega^2r/3$ to gravity in the hydrostatic equilibrium equation. The atmosphere is matched to a $T(\tau)$ law at an optical depth of $\tau=20$.

\subsection{Initial conditions}
 \label{Ss_IC}
 As mentioned in introduction, stellar evolution calculations require the definition of appropriate initial conditions, which set the initial AM. Stars with masses lower than about 2M$_{\odot}$ are born fully convective on the Hayashi track, and then develop a radiative zone on the Henyey track before they reach the zero-age-main-sequence and ignite nuclear reactions through the CNO cycle. There, they develop a convective core and the outer convective zone shrinks drastically in a mainly radiative envelope.

 This whole pre-main-sequence (PMS) stage for a stellar mass typical of $\gamma$ Dor stars lasts around 8 to 10 Myrs depending on the metallicity. The PMS is too short for internal transport of angular momentum by rotationally induced processes or stellar winds to slow down the star significantly. While contracting, these stars are prevented to spin up due to tight interaction with their residual accretion disk, before it fully dissipates after a few Myrs (up to around 5 Myrs for the low-mass end of \GD stars).

 The star-disk interaction in this early stage involves a series of complex mechanisms which include accretion and magnetic interaction between the star and the disk, and the issue of AM exchange between the star and its environment still remains controversial. The choice here is to rely on an empirical description of that interaction, which simply assumes that the stellar angular velocity remains constant as long as the star interacts with its disk \citep[see][]{Bouvier1997}. The problem is then reduced to two free parameters: the accretion disk lifetime $\tau_{\rm disk}$, i.e. the time during which the star is forced to co-rotate with its disk, and the period of rotation of the disk P$_{\rm disk}$. 

\section{Rotation distributions in young clusters}
\label{S_init}
In a similar fashion as in \cite{Amard2016}, or before them \cite{Gallet2013}, we aim at anchoring the evolution of AM at the pre-main-sequence stage by reproducing the rotational distributions found in very young stellar clusters. The specificity here is that we narrow down the mass range of stars in these clusters to the corresponding masses of \GD  stars, i.e. 1.3 to 1.9 M$_{\odot}$.

In order to set up the free parameters of the disk locking model (see Sect. \ref{Ss_IC}), we have selected stellar clusters younger than 20 Myrs, which have surface rotation measurements available in the literature, and which would contain a significant number of stars in our mass range (1.3 to 1.9 M$_{\odot}$). Three clusters fulfil these requirements: NGC2264, NGC2362, and hPer. The youngest, NGC2264 has an age around 3 Myrs \citep[][and references therein]{Affer2013}, its stellar surface rotation periods have been measured by rotational signature in the stellar light-curves by \cite{Venuti2017}. NGC2362 is about 5 Myrs old, and the rotational data have been taken from \cite{Irwin2008a}. Finally hPer (NGC869) is around 13 Myr old, and its stellar surface rotations have been measured by \cite{Moraux2013}. We have computed the median, 20$^{\rm th}$ and 80$^{\rm th}$ percentiles of these distributions. These have been taken respectively as reference points for initial conditions of typical models of 1.4, 1.6 and 1.8 M$_{\odot}$ (computed as described in Sect. \ref{S_model}). In other words, $\tau_{\rm disk}$ and P$_{\rm disk}$ have been tuned in order to fit these reference points in NGC2264 and NGC2362, and we have ensured that the models for the lowest mass also agree with the rotational distribution in hPer, when the disk has dissipated. 
\begin{figure}
  \centering
  	\includegraphics[width=0.9\linewidth]{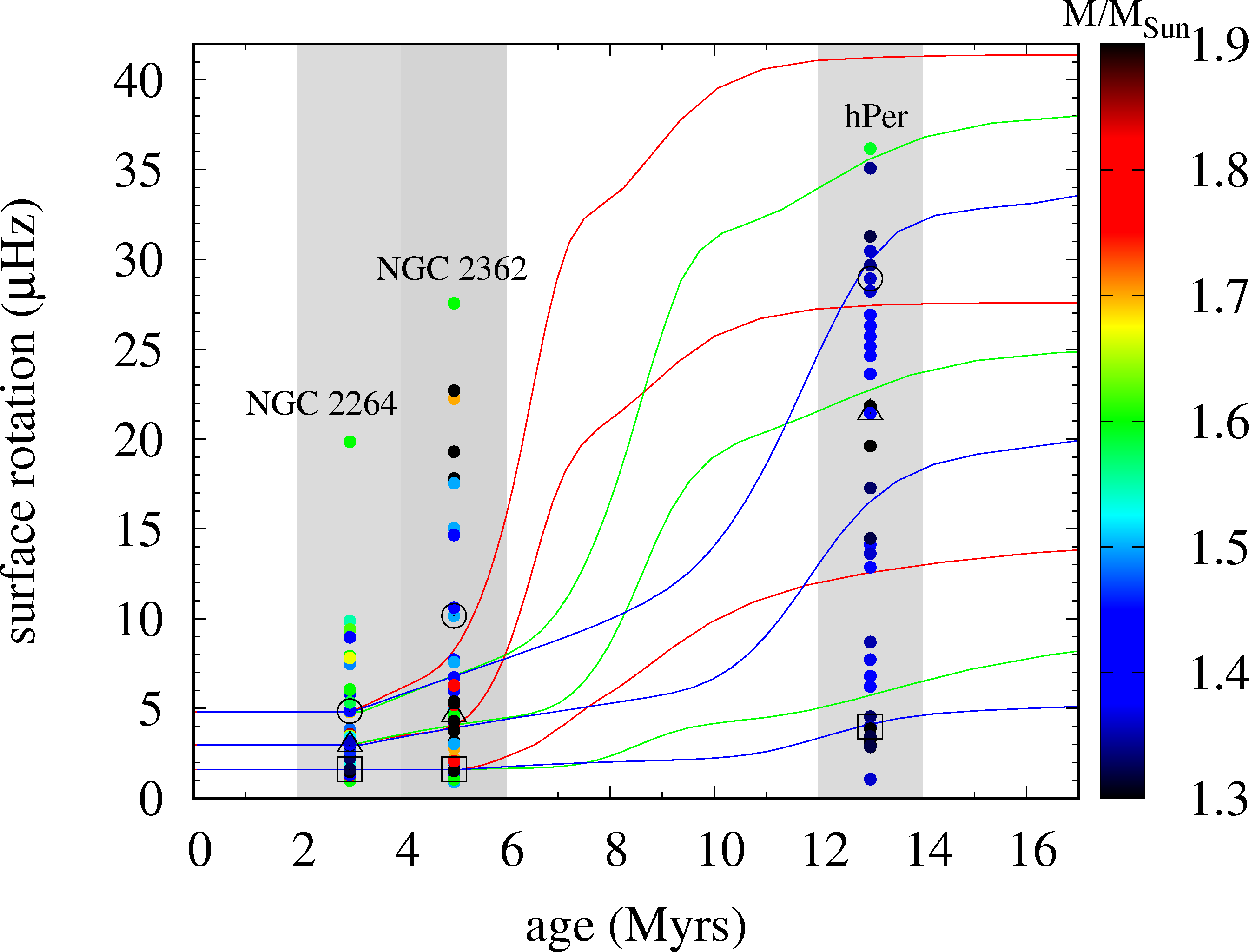}
        \caption{\label{fig:rot-vs-age_PMS} Surface rotation distributions in the three selected clusters NGC2264, NGC2362 and hPer, as a function of age. The colour codes for the stellar mass (from 1.3 to 1.9 M$_{\odot}$). The solid lines are given by evolutionary models of 1.4 (in blue), 1.6 (in green) and 1.8 M$_{\odot}$ (in red). The grey coloured areas symbolize the uncertainty on the age of the clusters which is of around 1 Myr. The squares, the triangles and the open circles represent the 20$^{\rm th}$ percentiles, the median, and the 80$^{\rm th}$ percentile of the rotation distributions in each cluster, respectively.}
 \end{figure}

The result is shown in Fig. \ref{fig:rot-vs-age_PMS}. The three sets of disk-locking parameters which allow to reproduce the distributions satisfactorily are:
\begin{itemize}
\item[-] P$_{\rm disk}$ = 2.4 days and $\tau_{\rm disk}$ = 3 Myrs,
\item[-] P$_{\rm disk}$ = 3.9 days and $\tau_{\rm disk}$ = 3 Myrs,
\item[-] P$_{\rm disk}$ = 7.2 days and $\tau_{\rm disk}$ = 5 Myrs .
\end{itemize}
The aim here is not to reproduce precisely the observed rotational distributions in these clusters, but to rather get the overall orders of magnitude correct. One should bear in mind that the uncertainties on age can be relatively important for these clusters ages. Indeed, the observed ages (data points: filled circles in Fig. \ref{fig:rot-vs-age_PMS}) are obtained by isochrone fitting, and therefore are contaminated by inaccuracies in the stellar evolution models used to generate the isochrones. Moreover, dispersion in age can come from different generation of stars in the stellar forming region. Finally, concerning the stellar models which are fitted to these data points (solid lines in Fig. \ref{fig:rot-vs-age_PMS}), they are also altered by the uncertainties on the birth lines location compared to what has been taken as age zero in the stellar evolution. 

While the distributions in NGC2264 and NGC2362 are correctly reproduced by the three sets of models, only the 1.4 M$_{\odot}$ models manage to fit the rotation distribution in hPer. Unfortunately, given that hPer only contains stars with masses around 1.3-1.5 M$_{\odot}$, it does not allow to constrain the higher mass models. At these very young age, we were not able to find clusters containing a significant number of stars of higher masses.
\begin{figure*}
   \hspace*{0.7cm}\includegraphics[width=0.36\linewidth]{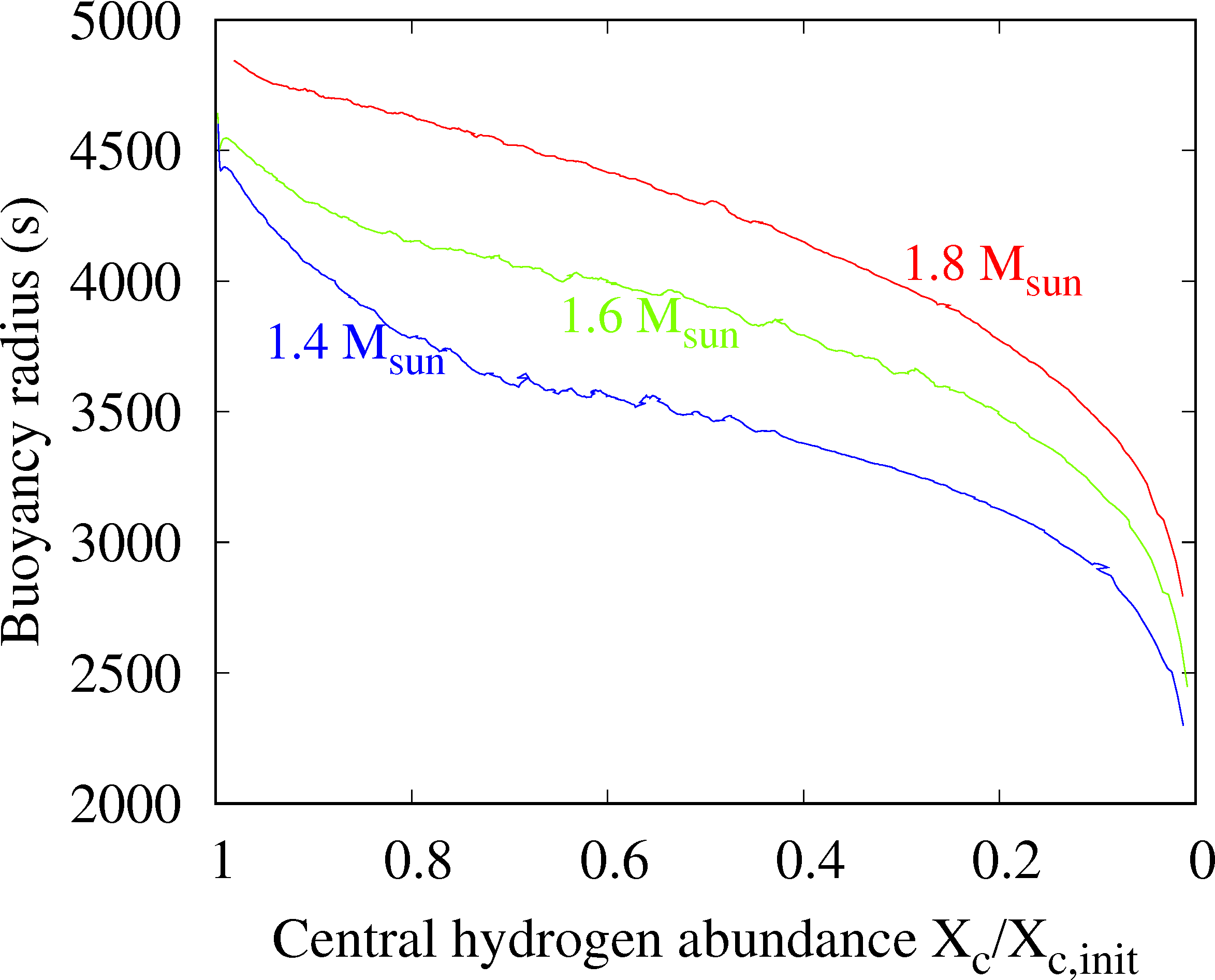}\hspace*{0.5cm}~\,~\, \hspace*{1.5cm}\, 
   \includegraphics[width=0.45\linewidth]{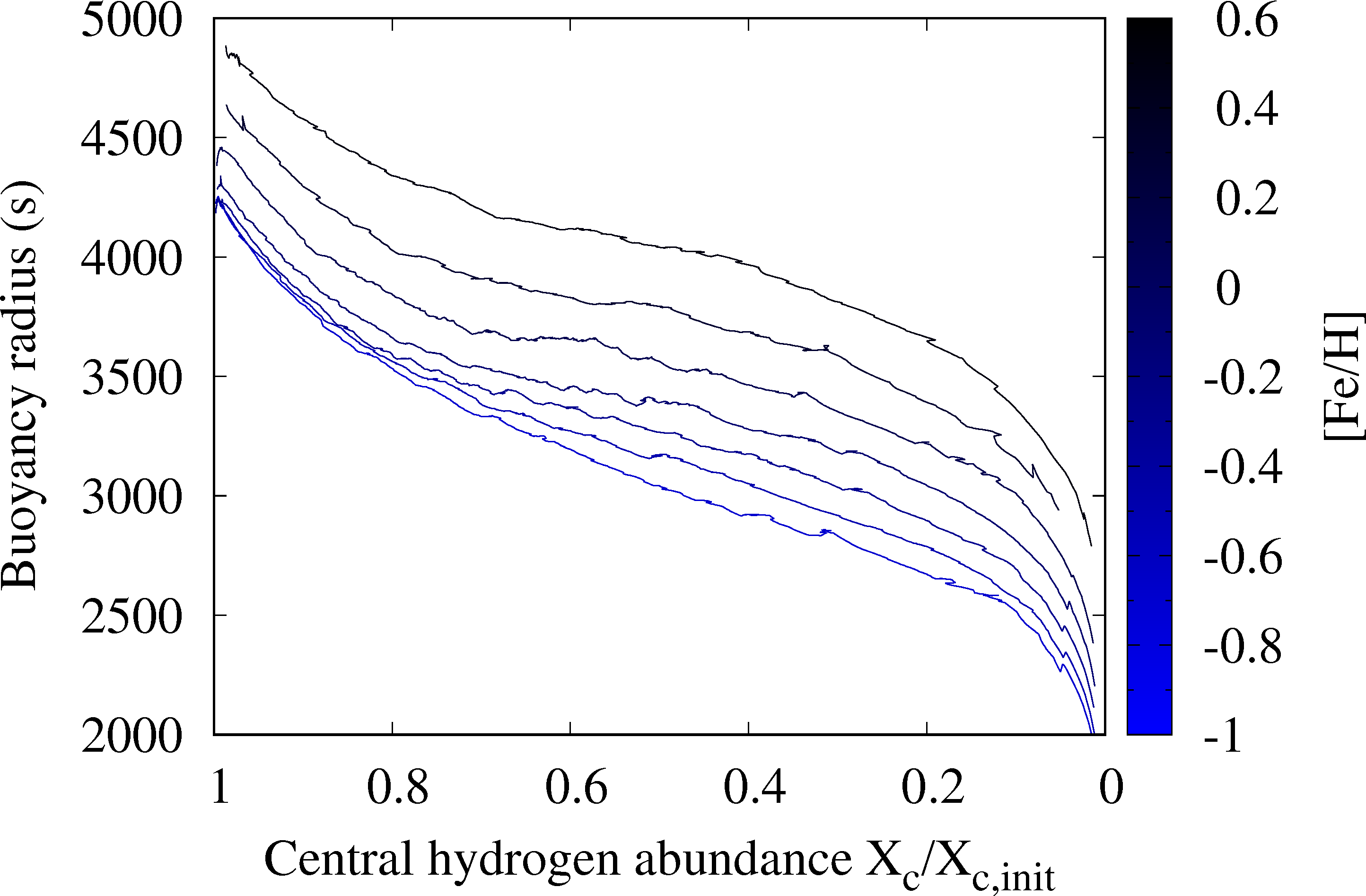}
   \caption{Buoyancy radius as a function of central hydrogen abundance X$_c$ (relative to the initial X$_c$) along main-sequence evolution. {\bf Left:} for three different masses typical of \GD \string : 1.4 in blue, 1.6 in green and 1.8 M$_{\odot}$ in red, same metallicity $\rm [M/H] = 0.1$. The colour coding is the same as in Fig\ref{fig:rot-vs-age_PMS}. {\bf Right:} for a 1.4 M$_{\odot}$ model, with varying metallicities from $\rm [M/H]$=-0.9 to 0.5, with increments of 0.2 dex. The X$_{\rm c} /$ X$_{\rm c,init} $ axis has been inverted to emphasize the behaviour with evolution. The models were computed with disk locking during 5 Myrs, to a disk rotating with a period of 7.2 days.}
   \label{fig:Buoy-vs-Xc_MS_TZ97}
 \end{figure*}

 \section{Buoyancy radius as an age indicator}
 \label{S_BR}
 For field stars on the main sequence, it is even less straightforward to determine the age with accuracy. \cite{Saio2015} mention the possibility to use the period spacing as a constraint of the evolution stage. Given the dependency of the period spacing to the rotation rate in the cavity probed by the g-modes, here we show that the buoyancy radius can be used as an age indicator for \GD  stars on the main sequence.

 \subsection{Behaviour of the buoyancy radius $P_0$}
 \label{Ss_BR}
 
The reason for this is that the buoyancy radius, $P_0$, of a star is a monotonously decreasing function of age. $P_0$ depends on the internal structure only, via N, the Brunt-V\"ais\"al\"a frequency, and can be expressed as:
  \begin{eqnarray}
    P_0 = 2 \pi^2 \left( \int_{\rm gc} \frac{N(r)}{r} dr \right)^{-1}, \label{eq:P0} \\
    \hbox{where }\hspace*{0.2cm} N(r)^2 = g \left( \frac{1}{\Gamma_1 P}\frac{dP}{dr} -  \frac{1}{\rho}\frac{d\rho}{dr} \right)\, \simeq \, \frac{g^2 \rho}{P} \left( \nabla_{ad} - \nabla + \nabla_{\mu} \right).
  \end{eqnarray}
Where the Brunt-V\"ais\"al\"a frequency is expressed by the help of g, the gravitational acceleration, $P$, the local pressure and $\Gamma_1$ the adiabatic exponent. $\nabla = d \ln T / d \ln P$ is the thermal gradient, $\nabla_{\rm ad}$ the adiabatic gradient, and $\nabla_{\mu} = d \ln \mu / d \ln P$, where $\mu$ is the mean molecular weight, . This last term shows the explicit contribution of the mean molecular weight gradient to the Brunt-Vaïssälä frequency, and hence to the buoyancy radius. The integral in (Eq. \ref{eq:P0}) is computed over the g-modes cavity. We will get back to this in Sect. \ref{Ss_meas}.

In the convective core, the efficient convection induces an almost adiabatic stratification and mixes the material, $N^2$ is therefore equal to 0. Above the convective core, the $N^2$ profile depends on whether the star is on a core-contracting or core-expanding main sequence. For the physics we have considered, above around 1.3 M$_{\odot}$, the convective core retreats along the main sequence, and leaves behind, in the radiative zone, a region of strong molecular weight gradient $\nabla \mu$. The buoyancy radius strongly depends on both the extent of the convective core, and the chemical stratification above the core. As a result, it is a good indicator of evolution on the main sequence of $\gamma$ Dor stars. The evolution of a star's $P_0$  is then straightforward to understand: along the main sequence evolution, the chemical abundance contrast increases at the edge of the core, thereby increasing the mean molecular weight gradient $\nabla \mu$, and the convective core contracts, therefore increasing the interval where $N^2$ is non-zero, which induces a monotonously increasing peak in the Brunt-V\"aiss\"al\"a frequency.

  The behaviour of $P_0$ is illustrated in Fig. \ref{fig:Buoy-vs-Xc_MS_TZ97}, where it is plotted against the central hydrogen abundance, for models with varying masses (left) or varying metallicities (right). As expected, both a higher metallicity and a higher mass result in larger convective cores, and therefore a higher buoyancy radius at a given age. This degeneracy of mass and metallicity has to be carefully accounted for when comparing models with observations (see Sect. \ref{S_results}). 
  
\subsection{Seismic measurement of the buoyancy radius and the near-core rotation}
\label{Ss_meas}
Conveniently, by making reasonable approximations, $P_0$ can be determined from \GD stars g-modes spectra. As they pulsate in high radial order g-modes, their pulsations are located in the asymptotic regime of g-modes. Without rotation, their spectra can be well approximated by the first order asymptotic theory of \cite{Tassoul1980}, which predicts that the periods of oscillation can be approximated at first order as:
 \begin{equation}
    P^{co}_{n,\ell,m} \, \simeq \, \frac{P_0 \left( n+\epsilon \right)}{\sqrt{\ell (\ell+1)}}, 
 \end{equation}
 where n is the radial order, $\ell$ the angular degree, $m$ the azimutal order, and $\epsilon$ is nearly constant.
 
  For \GD stars, which are moderate to fast rotators, rotation can be accounted for through the traditional approximation of rotation (TAR) to a certain extent. The TAR assumes that a star is spherically symmetric and that the latitudinal component of the rotation vector in the Coriolis force can be neglected. The first authors to apply the TAR in the stellar case were \cite{Berthomieu1978}, see \cite{Unno1989} for a complete mathematical derivation. The principle is that under such assumptions, and assuming solid body rotation, the equation system of stellar oscillations including rotation is separable in terms of a radial component and the Hough functions. The eigenvalues of the Hough functions are the $\lambda_{\ell,m}(s)$ functions which depend on the angular degree, the azimuthal order and the spin factor $s=2 P^{co}_{n,\ell,m}/P_{rot}$. In the asymptotic regime, the TAR allows to express the period of modes in the co-rotating frame, including the effect of rotation, at first order, as:
  \begin{equation}
P^{co}_{n,\ell,m} \simeq \frac{P_0 \left( n+\epsilon\right)}{\sqrt{\lambda_{\ell,m}(s)}}. 
  \end{equation}

  In Christophe et al. (in prep), the authors derived a new method to determine the rotation period $P_{rot}$ and the buoyancy radius $P_0$ from \Kepler observations of g-modes series in \GD stars. The principle is to use a set of trial rotation periods, that allows us to change from the inertial frame to the co-rotating frame of reference, and then stretch the periodogram by multiplying it by the corresponding $\sqrt{\lambda_{\ell,m}(s)}$ functions, and search for periodicities in the stretched periodogram by the mean of a direct Fourier transform. The systematic errors due to uncertainties in modes frequency determination are calculated by propagating them by means of a Monte-Carlo simulation. The authors also assess the biases induced by the use of the traditional approximation, by testing it against complete calculations performed with the {\sc acor} code \citep{Ouazzani2012b,Ouazzani2015}. These inaccuracies are difficult to assess directly for each measure, because they require the knowledge of the true stellar structure. They are therefore determined with the help of hare and hounds exercises on synthetic oscillation spectra --calculated with the non-perturbative method-- and amounts to maximum 8.2\% for the buoyancy radius and 16.8 \% for the rotation frequency in the worst case.

\begin{figure}
    \centering
    \hspace*{-0.3cm}\includegraphics[width=1.02\linewidth]{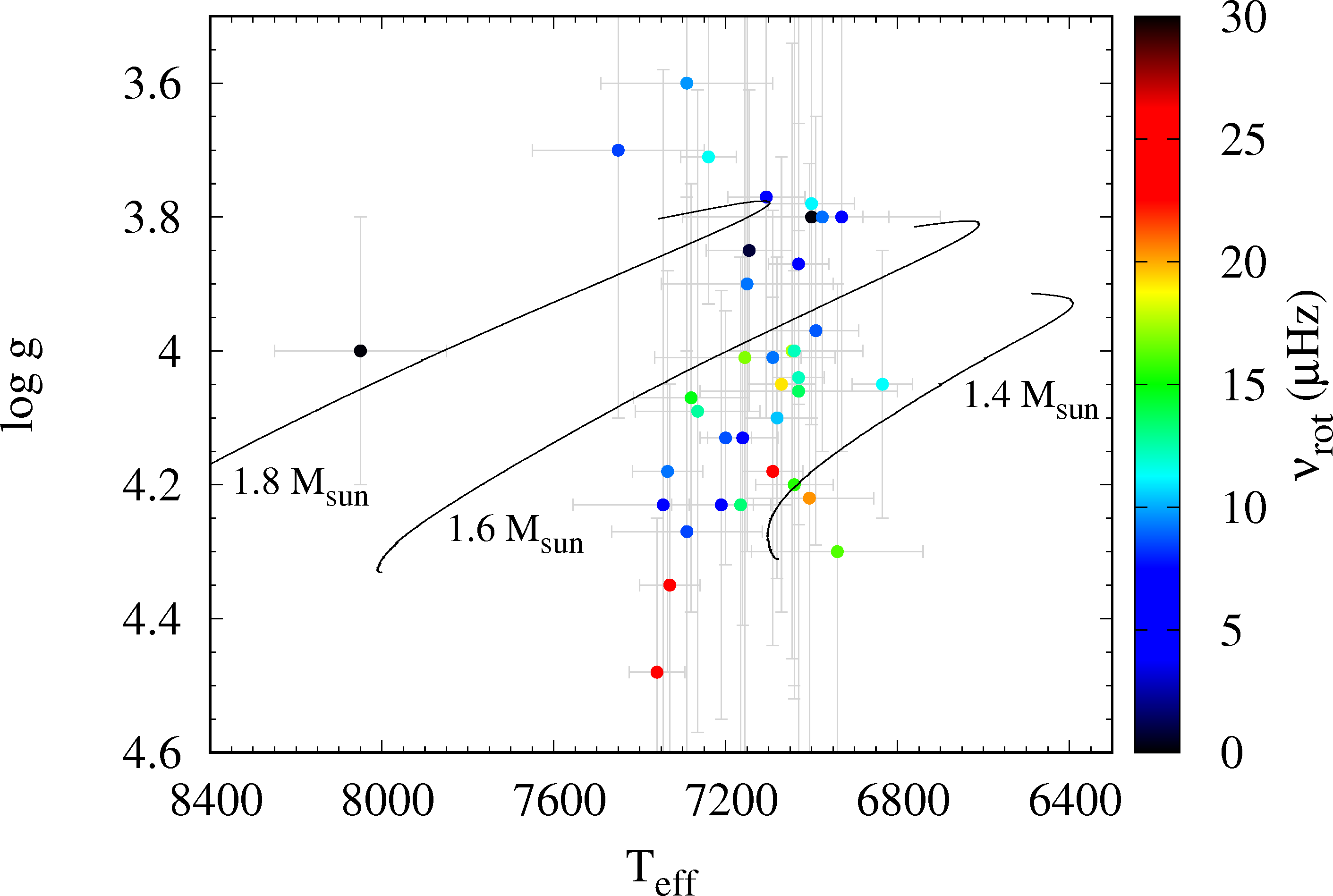}
    \caption{Kiel diagram with the observed stars sample (coloured circles). The atmospheric data use to build this diagram were taken from \cite{VanReeth2016} and the KASOC catalogue. The colour codes for their measured near-core rotation. To guide the eye, stellar models of 3 typical masses of \GD stars have been plotted in black solid lines. }
    \label{fig:Obs_HR}
  \end{figure}

  This method was used in order to determine the buoyancy radii and the rotation periods for the four stars previously presented in \cite{Ouazzani2017}. We complete the sample with a subset of 32 stars of the \cite{VanReeth2015} sample. Note that two stars previously modelled by \cite{Kurtz2014} and \cite{Saio2015} are included in the sample. The results are given in a Kiel diagram Fig.\ref{fig:Obs_HR}, where we show that the stars in the sample cover the whole \GD stars instability strip. The distribution in terms of near-core rotation frequency ($\nu_{\rm rot}$) seem to indicate that slow core-rotators lie all over the instability strip, while the faster ones occur more in the young-and-less-massive area of the diagram.\\

 \begin{figure}
    \centering
    \includegraphics[width=0.975\linewidth]{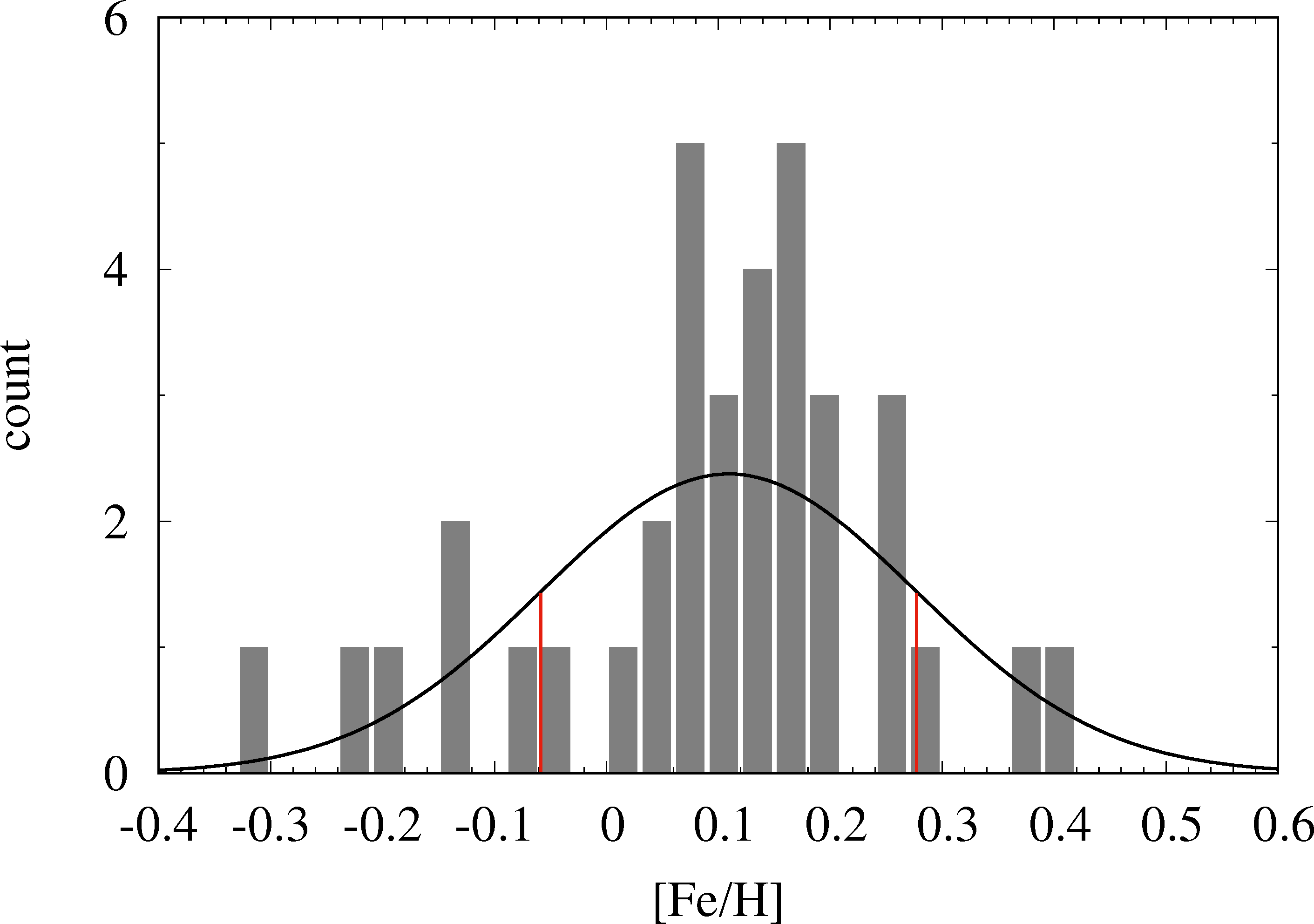}
    \caption{Histogram of the distribution of metallicity in the sample of stars. The black line is the Gaussian fit with a mean metallicity of $\overline{\left[\rm M/H\right]} = +0.11$ and a standard deviation $\sigma_{\left[\rm M/H\right]}=0.17$ dex. The red lines delimit the $\overline{\left[\rm M/H\right]}\pm \sigma_{\left[\rm M/H\right]}$ interval used for the computations of stellar models.}
    \label{fig:FeoH_histo}
  \end{figure}

  These measurements are then compared with evolutionary models, for which we have computed $P_0$ using the integral expression (Eq.\ref{eq:P0}). The rotation periods, on the one hand, are derived using the method of Christophe et al. (in prep) based on the TAR that considers the star as a rigid rotator. In principle, we assume that it is equivalent to considering a rotation period averaged over the cavity of the observed modes, given that among a series of high-order g-modes the cavity extent varies little. 

  On the other hand, the models result from evolutionary calculations including transport of AM by the processes defined in Sect.\ref{S_model}. Transport by shear induced turbulence and meridional circulation Z92  give a shellular rotation profile. Rather than a central rotation period, one should consider the near-core rotation period for comparison with observations. In order to calculate these near-core rotation periods, or angular velocities, we choose to take the average velocity over the g-modes cavity, weighted by the Brunt-V\"aiss\"al\"a frequency:
  \begin{equation}
    \label{eq:Om_N2}
    <\Omega> \, = \, \frac{\int_{\rm gc} \, \Omega(r) \, N(r)\, \frac{dr}{r}}{\int_{\rm gc}\, N(r)\, \frac{dr}{r}}.
  \end{equation}
  In order to determine the integral boundaries in Eq.(\ref{eq:P0}) and Eq. (\ref{eq:Om_N2}), the extent of g-modes cavity has to be determined in each of the numerous computed stellar models. Our concern was to do so, without having to calculate the oscillation spectra at each evolutionary time step. Non-adiabatic calculations in \GD stars \citep{Dupret2005,Bouabid2013} showed that the excited modes had their radial orders comprised between around -15 and -50. The lower boundary (-15) is set by the necessity of having the oscillation time-scale close to the convective turnover time-scale at the bottom of the convective envelope, while the higher boundary (-50) is set by the fact that high radial order modes are radiatively damped. Considering that the modes cavity vary little over the range of excited radial order, the g-modes cavity is determined as being the cavity for the $n=-25$ mode, which frequency is computed by the use of the \cite{Tassoul1980} asymptotic formula, and the effect of rotation being accounted for by a \cite{Ledoux1951} splitting. Then, the g-modes cavity is defined as the interval where the thereby determined $n=-25$ mode frequency is lower than the Brunt-V\"aiss\"al\"a frequency and the Lamb frequency.  We consider that this is close enough to an average on g-modes cavity, for a first attempt to compare with observations.

\subsection{Sample of observed stars}
\label{Ss_FeoH}

We have gathered a total of 37 \GD stars, which have been observed with {\it Kepler} during the nominal mission. Because we want global parameters determined as precisely as possible, we have built up from the sample of spectroscopically observed stars in \cite{VanReeth2015}, to which we added the four stars mentionned in \cite{Ouazzani2017}. For these last four stars, we have relied on the {\it Kepler} input catalogue for the atmospheric parameters. All the stars of the sample have gone through Christophe et al. algorithm for near-core rotation and buoyancy radius determinations. For four stars out of the 33 taken in \cite{VanReeth2015} sample, we were not able to reliably decipher g-modes series (due to more conservative selection criteria), so we used \cite{VanReeth2016} determinations of $\nu_{rot}$ and period spacings.

Before confronting models to observations, we have to handle the mass-metallicity degeneracy of the buoyancy radius. To do so, we explore the metallicity distribution of the sample of stars. For the four stars taken from \cite{Ouazzani2017}, the metallicity was taken from the {\it Kepler} input catalogue, while for the rest of the stars, the values come from \cite{VanReeth2015}. The metallicites range from $\left[\rm M/H\right]=-0.03$ to $+0.25$ with uncertainties reaching $\pm 0.30$ dex. As illustrated in Fig.\ref{fig:FeoH_histo}, they follow an overall close-to-Gaussian distribution, centred around $\overline{\left[\rm M/H\right]}=+0.11$ with standard deviation of $\sigma_{\left[\rm M/H\right]}=0.17$ dex. The strategy here is to compute buoyancy radii along evolution for each of the stellar models masses we selected, and corresponding to the two metallicities: $\overline{\left[\rm M/H\right]}-\sigma_{\left[\rm M/H\right]}$ and $\overline{\left[\rm M/H\right]}+\sigma_{\left[\rm M/H\right]}$.

  \section{Results}
  \label{S_results}
 \begin{figure}
  	\includegraphics[width=1.\linewidth]{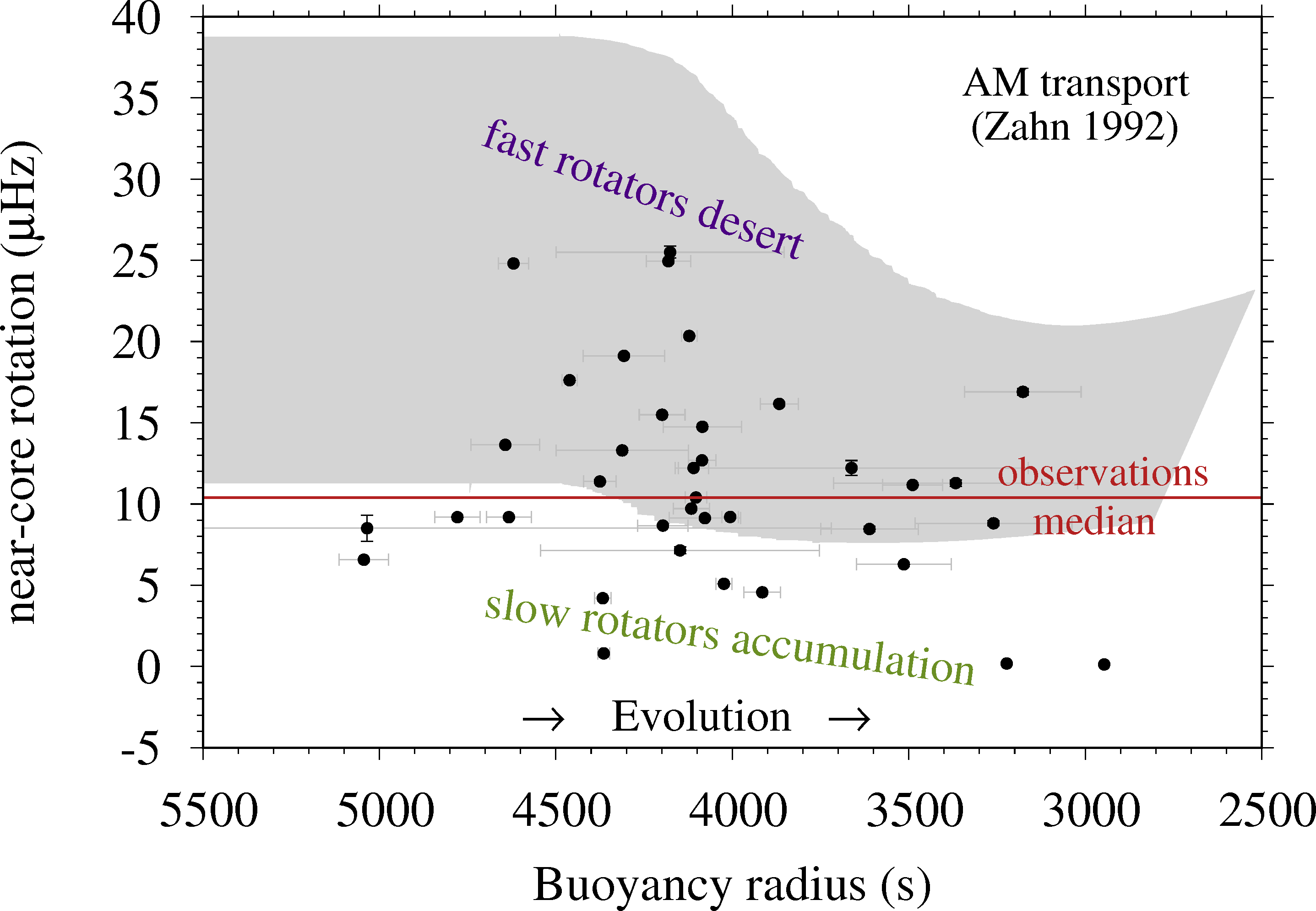}
        \caption{Measured near-core rotation rates as a function of the buoyancy radius (black circles), error bars are taken from \cite{VanReeth2016}. The grey area represents the interval of possible rotation rates given the initial conditions set in Sect.\ref{S_init}, and for the transport of angular momentum as formalized by Z92. The red solid line gives the median of the rotation distribution in the main sequence observed sample.}
    \label{fig:rot-vs-age_MS_TZ97_TA}
\end{figure}

\subsection{Rotational transport of angular momentum}
\label{Ss_ResultsZ92}

\begin{figure*}[p]
  \centering
  	\includegraphics[width=1\linewidth]{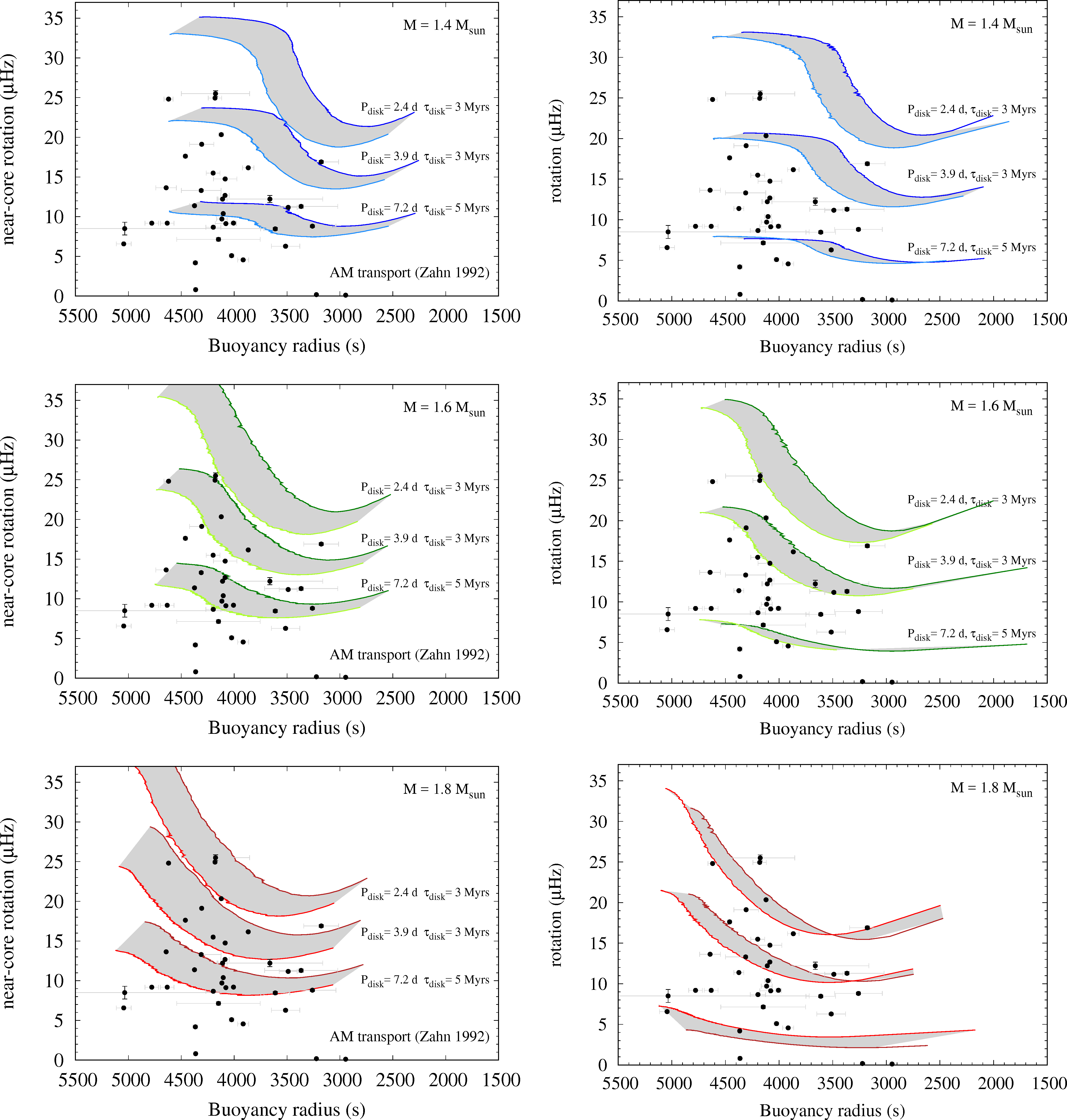}
        \caption{Rotational evolution as a function of the buoyancy radius. We illustrate the observations in black filled circles. The grey areas represent the rotation rates versus buoyancy radius covered when the metalicity varies from $\overline{\left[\rm M/H\right]} - \sigma_{\left[\rm M/H\right]}$ to $\overline{\left[\rm M/H\right]} + \sigma_{\left[\rm M/H\right]}$, where the mean and the standard deviation of the observed sample are considered. The colored solid lines are the evolutionary models for three typical masses of \GD stars \string: 1.4 M$_{\odot}$ in blue (top), 1.6 M$_{\odot}$ in green (middle) and 1.8 M$_{\odot}$ in red (bottom). In each panel, the three sets of models relate to the different initial conditions set up in Sect.\ref{S_init}: from top to bottom\string : disk locking at a period of 2.4 days during 3 Myrs, 3.9 days during 3 Myrs, and 7.2 days during 5 Myrs. {\bf Left\string:} The evolution of the angular momentum distribution has been calculated with transport by meridional circulation and shears (Z92). {\bf Right\string:} Solid-body rotation is enforced all along evolution. The x-axis has been inverted so that the behavior with evolution goes from left to right. }
    \label{fig:rot-vs-BR}
 \end{figure*}

We compare the observations to the evolutionary models including angular momentum transport as described in Sect.\ref{S_model} in a near-core rotation versus buoyancy radius plane. The results are given in Fig. \ref{fig:rot-vs-BR}. Note that stars with detectable binary companions were excluded from the sample. We compared rotational evolution on the main sequence for three typical masses of \GD stars: 1.4M$_{\odot}$ (top), 1.6M$_{\odot}$ (middle) and 1.8 M$_{\odot}$ (bottom), each time considering two metallicities: $\left[\rm M/H\right] = -0.06$ and $\left[\rm M/H\right] = +0.28$. In Fig.\ref{fig:rot-vs-BR} (left) the evolutionary models account for transport of angular momentum via meridional circulation and shear-induced turbulence according to the formalism of Z92, as explained in Sect.\ref{S_model}.

The measured rotation frequencies are spread over a large range: from 0.8$\mu$Hz for KIC 9751996, to 26.1$\mu$Hz for KIC 7365537. But in terms of buoyancy radii, they only cover the higher half of the interval covered by the models. However, this can be explained by exploring the evolutionary tracks: the lower half of the buoyancy radius interval corresponds to the {\it second contraction} evolutionary stage, when the stars have exhausted their hydrogen reservoir, and gravitational contraction takes over. In other words, this stage lasts a very short amount of time at the end of the main sequence. For instance for a 1.8M$_{\odot}$ model, this represents around 30 Myrs, i.e. 2.5 \% of the total main-sequence duration. Statistically, there is much less chance to observe a star in that stage, than in the main sequence.

On each panel of Fig.\ref{fig:rot-vs-BR}, for each mass and initial condition, we have plotted two curves, corresponding to the $\pm 1 \sigma$ of the metallicity distribution. The lower the metallicity, the higher the buoyancy radius is. Indeed, a lower metallicity induces a lower opacity, which impacts the radiative gradient, and therefore the limit of the convective core. The resulting increase of the extent of the convective core,  generates a decrease of the integral in (Eq.\ref{eq:P0}), leading to an increase of the buoyancy radius.

  Compared to the models, the measured rotation rates seem to be globally shifted towards lower rotation (Fig.\ref{fig:rot-vs-BR}). In the sample studied here, there is no observed $\gamma$ Dor star with rotation frequency higher than about $\sim$26 $\mu$Hz, while all the models corresponding to the fast initial conditions (P$_{\rm disk}$ = 2.4 days and $\tau_{\rm disk}$ = 3 Myrs) give such rotation values or higher. The stars which PMS progenitors were locked to their disk at 2.4 days are either not observed, or seem to be braked by a missing mechanism. Even for the low-mass stellar models, with a distribution of rotation periods constrained by hPer observations at 13 Myrs, do not fit the observations later, on the main sequence. 

  On the low-rotation end, even with the slowest initial conditions, the models cannot fit the observed slow rotators: 15 stars out of 37, i.e. 41\%,  fall out of the rotation interval covered by the models. 68\% of the targets have metallicities within the $\pm 1 \sigma$ interval. Therefore the spread in metallicity cannot explain the discrepancy between the rotation interval covered by the models and the observations.

\subsection{Enforcing rigid rotation}
 \label{Ss:SB}
  We have tested a different angular momentum transport hypothesis: solid body rotation all along evolution. This is not physically motivated, but it can be considered as a limiting case of instantaneous and highly efficient transport inside the star. The corresponding evolutionary models are confronted to observations in Fig. \ref{fig:rot-vs-BR} (right). Here again, very few observed stars (and only for the models with M=1.8M$_{\odot}$) fall in the area corresponding to the fastest initial conditions. However, the comparison is more favourable than in the case of hydrodynamical angular momentum transport (Fig. \ref{fig:rot-vs-BR}, left). The real difference between the results obtained for the two transport prescriptions is at the low-rotation frequency end. The slow rotators (around $\sim 5 \mu$Hz) are much better reproduced by the models relying on the solid body assumption.

This suggests the existence of a mechanism transporting angular momentum from the core to the outer part of the star. In this study, we assumed that these stars cannot sustain generation of magnetic field through a solar-like dynamo process, therefore models do not loose AM at the surface. As a result, in the presented models, the AM transport mechanism rigidifies the rotation profile. However, the absence of constraint on the envelope rotation in these stars prevents any strong conclusions on the profile per say.

In the next subsections, we estimate the efficiency of the missing mechanism in a similar spirit as in \cite{Eggenberger2017} for the red-giant phase, in order to characterize the missing AM transport processes.

 \subsection{Uncertainties on the horizontal turbulent viscosity}
 \label{Ss:Dh}
We first attempt to extract AM from the core while remaining within the Z92 framework, by exploring the uncertainties in turbulent viscosity coefficients. It has been long known that large uncertainties remain concerning the turbulent coefficients in the formalism \citep[see for instance][and references therein]{Mathis2004b}. 
In particular, the prescriptions for the horizontal coefficient of viscosity, $\nu_h$, can differ by as much as two orders of magnitude. The results given in Sect. \ref{Ss_ResultsZ92} are obtained with the \cite{Mathis2004b} prescription. Here we choose to multiply $\nu_h$ by a factor of $10^2$.  Such an increase of $\nu_h$ results in an enhancement of the meridional circulation by ensuring attenuation of the horizontal variation of the mean molecular weight \citep{Marques2013}. While it indirectly impacts the transport of chemical elements, the effect on the chemical stratification is marginal. The main effect is the enhancement of advective transport of AM through the meridional circulation.

\begin{figure}
    \centering
    \includegraphics[width=1\linewidth]{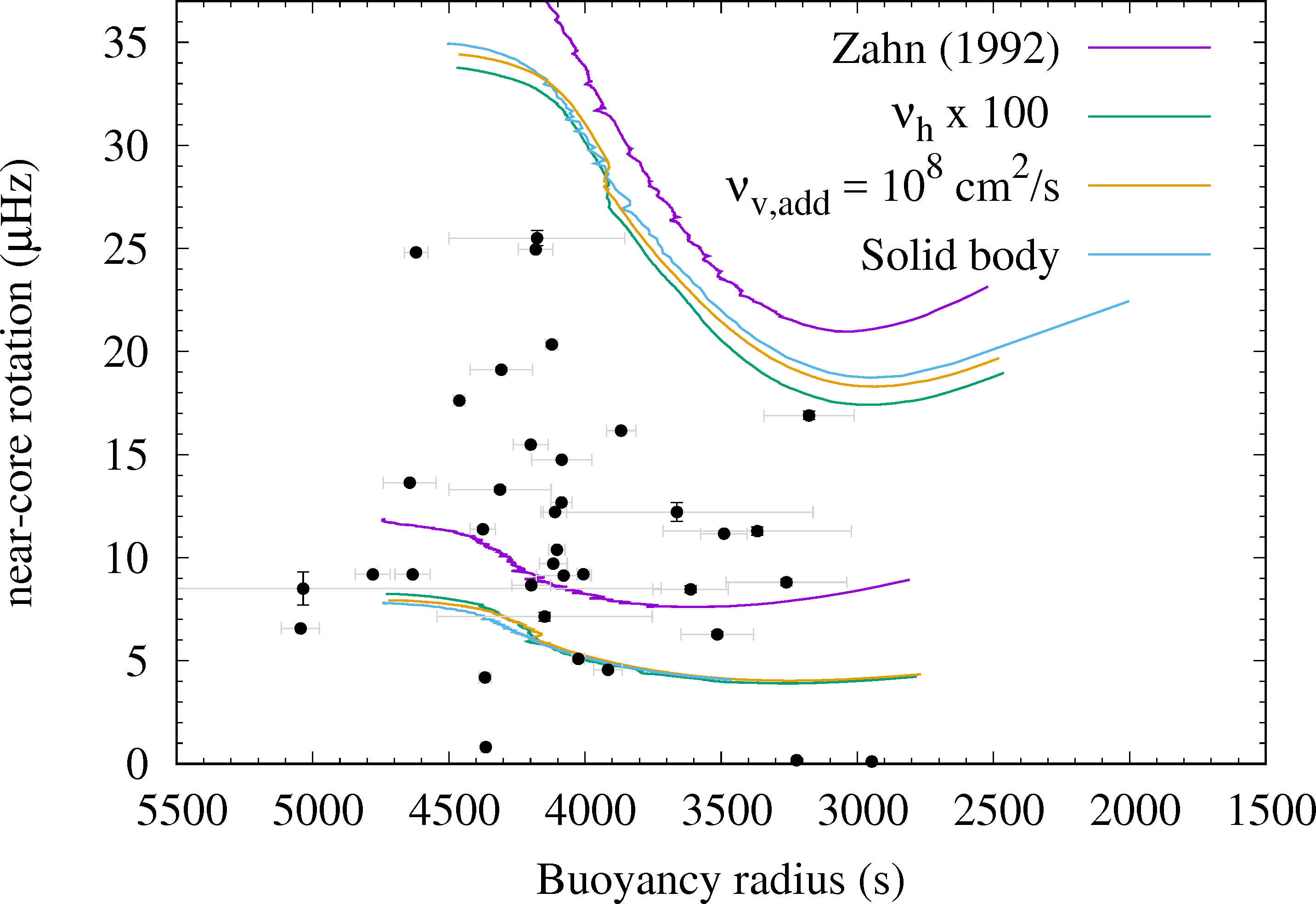}
    \caption{Near core rotation rates as a function of the buoyancy radius. The observed values are given by the black circles. The solid lines stand for models of angular momentum transport with different prescription: Z92 in purple (cf Sect.\ref{Ss_ResultsZ92}), with enhanced horizontal viscosity in green (cf Sect.\ref{Ss:Dh}), with enhanced vertical viscosity in yellow (cf Sect.\ref{Ss:Dv}), and with solid body enforced along evolution in blue (cf Sect.\ref{Ss:SB}). These models have been computed for two models of 1.6 M$_{\odot}$, one (top) corresponding to the fast rotation initial conditions given in Sect.\ref{S_init}, and for the lowest metallicity, and the other one (bottom) to slow initial conditions and high metallicity.}
    \label{fig:missingJ}
  \end{figure}

The results of such calculations are given in Fig. \ref{fig:missingJ} (green solid lines), by comparison with the unchanged Z92 models (purple solid lines), and the solidly rotating models (blue solid lines). This figure shows that amplifying $\nu_h$ by two orders of magnitude radically increases the core-to-envelope coupling to such an extent  that it is equivalent to simply enforcing solid body rotation all along evolution.

The two sets of lines in Fig.\ref{fig:missingJ} correspond to two combinations of metallicity and initial conditions for rotation: $[\rm M/H]=-0.06$, $\tau_{\rm disk}=3$Myrs, $\rm P_{\rm disk}=2.4$ days for the upper one, and $[\rm M/H]=+0.028$, $\tau_{\rm disk}=5$Myrs, $\rm P_{\rm disk}=7.2$ days for the lower one. They delimit the area where we expect to find at least 68\% of the observations if the prescriptions for transport of AM are correct. Figure \ref{fig:missingJ} shows that there is a clear improvement at the lower rotation end of the diagram, where an increased $\nu_h$ allows a much better agreement with observations. Indeed, 80\% of the measured rotation values lie in the possible {\it rotation versus Buoyancy radius} area.  

However, at the high rotation end, such an increased transport does not allow to slow down the models with the fast initial conditions. 

\subsection{Additional vertical diffusion of AM}
\label{Ss:Dv}
The rotational transport of angular momentum, as formalized by Z92 does not seem to satisfactorily reproduce the observed core rotation in $\gamma$ Doradus stars, particularly in the fast rotation regime. Hence, additional AM transport is suspected to operate. Because the physical nature of the missing mechanism remains unclear, we choose to model its impact as an additional vertically diffusive process, which efficiency, function of the vertical viscosity coefficient $\nu_{v, add}$, has to be higher than provided by the {\it standard} Z92 processes. In the transport equation, it consists of adding $\nu_{v, add}$ to $\nu_{v}$ in the second term of the right hand side of Eq.\ref{eq:AMT}, without modifying $D_v$ in Eq.\ref{eq:chem}.

Different values for $\nu_{v,add}$ are tested. For this additional viscosity to have an impact on the internal transport, it has to be stronger than $\nu_v$. Hence the process starts to be efficient for approximately $1\times 10^6 \rm cm^2.s^{-1}$. Higher values of $\nu_{v, add}$ are then tested as high as $1\times 10^8 \rm cm^2.s^{-1}$, the viscosity for which the evolution of the near core rotation overlaps with the one obtained assuming rigid rotation. This is illustrated Fig.\ref{fig:missingJ} in yellow solid lines.

The value of $\nu_{v,add}$ which allows to reproduce more than 64\% of the observed rotations is found to be approximately $5\times 10^6 \rm cm^2.s^{-1}$. In other words, in this case as in the previous subsection, this additional process allows to spin down the core rotation for the models corresponding to slow initial conditions sufficiently to reproduce the observed slowest rotators satisfactorily. 

However, as with horizontal viscosity (Sect.\ref{Ss:Dh}), the models with fast initial conditions are not slowed down enough to be shifted into the observed regime.




 \section{Discussion and conclusions}
 \label{S_cp}

 We have measured near-core rotation rates of a sample of 37 $\gamma$ Doradus stars on the main sequence, in order to test angular momentum transport models for intermediate-mass stars. We calculated rotational evolution of stellar models with masses typical of these stars from the PMS to the TAMS. To set the initial conditions for such evolution, we constrained simple disk-locking models to rotation distribution in young stellar clusters.

After being released from the disk, the angular momentum evolution of the stellar model is then dictated by the prescription for transport. The aim of this study is to use measurements of near core-rotation in \GD stars in order to constrain such prescriptions. 

Before doing so, confrontation of evolutionary calculations to observations require to define an evolution indicator. To that end, we have explored the dependence of the buoyancy radius $\rm P_0$ (Eq.\ref{eq:P0}) on stellar parameters, and showed that $\rm P_0$ could play that role, provided the degeneracy in mass and metallicity is carefully handled.

The angular momentum transport by meridional circulation and shear-induced turbulence (as described by Z92) was first explored. Fig.\ref{fig:rot-vs-age_MS_TZ97_TA} summarizes the results obtained: such models do not allow to reproduce the near-core rotation measured by g-modes period spacing in our sample of \GD stars. Not only does it fail to reach the particularly slow cores of 15 stars of the sample (slow rotators accumulation Fig.\ref{fig:rot-vs-age_MS_TZ97_TA}), but it also fails at slowing down models of stars which progenitors are the 20\% of fast rotators observed in young stellar clusters (fast rotators desert Fig.\ref{fig:rot-vs-age_MS_TZ97_TA}).
Considering these arguments, and as for the Sun and red giant stars, it is safe to assert that the transport of angular momentum as formalized by Z92 cannot explain the measurements of near-core rotation in main-sequence intermediate-mass stars we have at hand.

Furthermore, we compared the rotation measurements to models where solid-body rotation is enforced accross evolution, mimicking the effect of an efficient, yet to be identified mechansism. But this hypothesis allows to slow down the models corresponding to slow initial conditions so that more than 80\% of the observed rotation velocites are recovered.
The good agreement between these models and the observations, in the low rotation end, show that, as for red giant stars, in intermediate-mass stars on the main sequence, we also need a mechanism which efficiently transports angular momentum from the core of the star outward. 
However, the impact on the models evolved from fast initial conditions is marginal and does not allow to explain the lack of observations in the high rotation regime, the so-called fast rotators desert (Fig.\ref{fig:rot-vs-age_MS_TZ97_TA}). 
Whether the thereby extracted angular momentum is preserved in the envelope, i.e. results in rigid rotation, or ejected outside the star through winds is not clear, and the data at hand cannot allow us to solve this issue.

Trying to characterize further such AM extraction mechanism, we have shown that a rotational evolution similar to the one with solid body rotation could be obtained in two different ways. First, remaining in the framework of Z92, such a transport can be mimicked by advective AM transport through meridional circulation enhanced by additional horizontal viscosity ($\nu_h \times 100$).
Second, extraction of AM to that extent can be reached by adding vertical viscous transport with a viscosity of $\nu_{v,add} = 1 \times 10^8 \rm cm^2.s^{-1}$. Although both transport processes allow to reach the slow core rotations observed in \GD stars, they cannot explain the lack of measurement in the high rotation regime.

These results must be taken with two words of caution. Firstly, the observed sample used for comparison with models here is mainly composed of {\it Kepler} \GD stars in the \cite{VanReeth2016} sample. The distribution of projected rotation velocities in this sample can reach up to about 200 km.s$^{-1}$ with an average around $v \sin i =$ 60 km.s$^{-1}$. \cite{Royer2009} compiled a large sample of stars with projected rotational velocities from the literature, and sorted them by spectral type. They found that for the spectral types covered by the know \GD stars (A3 to F3), the average velocities per spectral type are between 120 km.s$^{-1}$ for the stars of spectral type A3, to 50 km.s$^{-1}$ for the stars of spectral type F3. This shows that the sample of stars used in the current study might suffer selection biases, and should be further investigated for completeness. 

Moreover, the models used in this study did not include convective overshooting above the core. Further investigation is needed to achieve an accurate physical description of this phenomenon, and its interaction with turbulent diffusion, and meridional circulation. However, it is expected that overshooting extends the g-modes cavity further away from the core. As a consequence, it results in an increase of the buoyancy radius of the models, which would give a better agreement with the observed values of $P_0$. 

However, there seem to be a lack of AM transport process in the evolutionary models presented here, which is at work during the pre-main-sequence phase, and which can be efficient enough to slow down the core of intermediate-mass stars on very short timescales before the stars reach the ZAMS. 
A decent candidate process could be transport by internal gravity waves (IGW). Refining the developments of \cite{Talon2005}, \cite{Charbonnel2013} investigated the effect of the transport by IGW generated from both convective zones: from the core (for models with masses higher than about 1.6 M$_{\odot}$) and from the envelope of PMS low and intermediate-mass stars. They have shown that IGW could drastically reshape the internal rotation profile of these stars, in particular causing a spin down of the central rotation, close to the ZAMS. Therefore, in absence of a dynamo-generated magnetic field, transport by IGW, particularly generated by the convective core, should be explored as a possible explanation for the observed fast rotators desert (Fig.\ref{fig:rot-vs-age_MS_TZ97_TA}).

Moreover, while enforcing the core spin down, such a mechanism would not necessarily induce rigid rotation. Hence, on the asteroseismic side, additional constraints on the rotation radial differential rotation would allow to put stronger contraints on the missing mechanism.

\begin{acknowledgements}
This research was supported by the CNES. The authors would like to thank Marc-Antoine Dupret and Morgan Deal for fruitful discussions.
\end{acknowledgements} 

\bibliographystyle{bibtex/aa}

\begin{thebibliography}{60}
\expandafter\ifx\csname natexlab\endcsname\relax\def\natexlab#1{#1}\fi

\bibitem[{{Abt} \& {Morrell}(1995)}]{Abt1995}
{Abt}, H.~A. \& {Morrell}, N.~I. 1995, \apjs, 99, 135

\bibitem[{{Affer} {et~al.}(2013){Affer}, {Micela}, {Favata}, {Flaccomio}, \&
  {Bouvier}}]{Affer2013}
{Affer}, L., {Micela}, G., {Favata}, F., {Flaccomio}, E., \& {Bouvier}, J.
  2013, \mnras, 430, 1433

\bibitem[{{Alexander} \& {Ferguson}(1994)}]{Alexander1994}
{Alexander}, D.~R. \& {Ferguson}, J.~W. 1994, \apj, 437, 879

\bibitem[{{Amard} {et~al.}(2016){Amard}, {Palacios}, {Charbonnel}, {Gallet}, \&
  {Bouvier}}]{Amard2016}
{Amard}, L., {Palacios}, A., {Charbonnel}, C., {Gallet}, F., \& {Bouvier}, J.
  2016, \aap, 587, A105

\bibitem[{{Angulo} {et~al.}(1999){Angulo}, {Arnould}, {Rayet}, {Descouvemont},
  {Baye}, {Leclercq-Willain}, {Coc}, {Barhoumi}, {Aguer}, {Rolfs}, {Kunz},
  {Hammer}, {Mayer}, {Paradellis}, {Kossionides}, {Chronidou}, {Spyrou},
  {degl'Innocenti}, {Fiorentini}, {Ricci}, {Zavatarelli}, {Providencia},
  {Wolters}, {Soares}, {Grama}, {Rahighi}, {Shotter}, \& {Lamehi
  Rachti}}]{Angulo1999}
{Angulo}, C., {Arnould}, M., {Rayet}, M., {et~al.} 1999, Nuclear Physics A,
  656, 3

\bibitem[{{Asplund} {et~al.}(2009){Asplund}, {Grevesse}, {Sauval}, \&
  {Scott}}]{Asplund2009}
{Asplund}, M., {Grevesse}, N., {Sauval}, A.~J., \& {Scott}, P. 2009, \araa, 47,
  481

\bibitem[{{Beck} {et~al.}(2012){Beck}, {Montalban}, {Kallinger}, {De Ridder},
  {Aerts}, {Garc{\'{\i}}a}, {Hekker}, {Dupret}, {Mosser}, {Eggenberger},
  {Stello}, {Elsworth}, {Frandsen}, {Carrier}, {Hillen}, {Gruberbauer},
  {Christensen-Dalsgaard}, {Miglio}, {Valentini}, {Bedding}, {Kjeldsen},
  {Girouard}, {Hall}, \& {Ibrahim}}]{Beck2012}
{Beck}, P.~G., {Montalban}, J., {Kallinger}, T., {et~al.} 2012, \nat, 481, 55

\bibitem[{{Belkacem} {et~al.}(2015){Belkacem}, {Marques}, {Goupil}, {Mosser},
  {Sonoi}, {Ouazzani}, {Dupret}, {Mathis}, \& {Grosjean}}]{Belkacem2015b}
{Belkacem}, K., {Marques}, J.~P., {Goupil}, M.~J., {et~al.} 2015, \aap, 579,
  A31

\bibitem[{{Berthomieu} {et~al.}(1978){Berthomieu}, {Gonczi}, {Graff},
  {Provost}, \& {Rocca}}]{Berthomieu1978}
{Berthomieu}, G., {Gonczi}, G., {Graff}, P., {Provost}, J., \& {Rocca}, A.
  1978, \aap, 70, 597

\bibitem[{{B{\"o}hm-Vitense}(1958)}]{Bohm-Vitense1958}
{B{\"o}hm-Vitense}, E. 1958, \zap, 46, 108

\bibitem[{{Bouabid} {et~al.}(2013){Bouabid}, {Dupret}, {Salmon},
  {Montalb{\'a}n}, {Miglio}, \& {Noels}}]{Bouabid2013}
{Bouabid}, M.-P., {Dupret}, M.-A., {Salmon}, S., {et~al.} 2013, \mnras, 429,
  2500

\bibitem[{{Bouvier} {et~al.}(1997){Bouvier}, {Forestini}, \&
  {Allain}}]{Bouvier1997}
{Bouvier}, J., {Forestini}, M., \& {Allain}, S. 1997, \aap, 326, 1023

\bibitem[{{Cantiello} {et~al.}(2014){Cantiello}, {Mankovich}, {Bildsten},
  {Christensen-Dalsgaard}, \& {Paxton}}]{Cantiello2014}
{Cantiello}, M., {Mankovich}, C., {Bildsten}, L., {Christensen-Dalsgaard}, J.,
  \& {Paxton}, B. 2014, \apj, 788, 93

\bibitem[{{Casagrande} {et~al.}(2007){Casagrande}, {Flynn}, {Portinari},
  {Girardi}, \& {Jimenez}}]{Casagrande2007}
{Casagrande}, L., {Flynn}, C., {Portinari}, L., {Girardi}, L., \& {Jimenez}, R.
  2007, \mnras, 382, 1516

\bibitem[{{Chaboyer} \& {Zahn}(1992)}]{Chaboyer1992}
{Chaboyer}, B. \& {Zahn}, J.-P. 1992, \aap, 253, 173

\bibitem[{{Charbonnel} {et~al.}(2013){Charbonnel}, {Decressin}, {Amard},
  {Palacios}, \& {Talon}}]{Charbonnel2013}
{Charbonnel}, C., {Decressin}, T., {Amard}, L., {Palacios}, A., \& {Talon}, S.
  2013, \aap, 554, A40

\bibitem[{{Deheuvels} {et~al.}(2012){Deheuvels}, {Garc{\'{\i}}a}, {Chaplin},
  {Basu}, {Antia}, {Appourchaux}, {Benomar}, {Davies}, {Elsworth}, {Gizon},
  {Goupil}, {Reese}, {Regulo}, {Schou}, {Stahn}, {Casagrande},
  {Christensen-Dalsgaard}, {Fischer}, {Hekker}, {Kjeldsen}, {Mathur}, {Mosser},
  {Pinsonneault}, {Valenti}, {Christiansen}, {Kinemuchi}, \&
  {Mullally}}]{Deheuvels2012}
{Deheuvels}, S., {Garc{\'{\i}}a}, R.~A., {Chaplin}, W.~J., {et~al.} 2012, \apj,
  756, 19

\bibitem[{{Dupret} {et~al.}(2005){Dupret}, {Grigahc{\`e}ne}, {Garrido},
  {Gabriel}, \& {Scuflaire}}]{Dupret2005}
{Dupret}, M.-A., {Grigahc{\`e}ne}, A., {Garrido}, R., {Gabriel}, M., \&
  {Scuflaire}, R. 2005, \aap, 435, 927

\bibitem[{{Eggenberger} {et~al.}(2017){Eggenberger}, {Lagarde}, {Miglio},
  {Montalb{\'a}n}, {Ekstr{\"o}m}, {Georgy}, {Meynet}, {Salmon}, {Ceillier},
  {Garc{\'{\i}}a}, {Mathis}, {Deheuvels}, {Maeder}, {den Hartogh}, \&
  {Hirschi}}]{Eggenberger2017}
{Eggenberger}, P., {Lagarde}, N., {Miglio}, A., {et~al.} 2017, \aap, 599, A18

\bibitem[{{Eggenberger} {et~al.}(2012){Eggenberger}, {Montalb{\'a}n}, \&
  {Miglio}}]{Eggenberger2012}
{Eggenberger}, P., {Montalb{\'a}n}, J., \& {Miglio}, A. 2012, \aap, 544, L4

\bibitem[{{Formicola} {et~al.}(2004){Formicola}, {Imbriani}, {Costantini},
  {Angulo}, {Bemmerer}, {Bonetti}, {Broggini}, {Corvisiero}, {Cruz},
  {Descouvemont}, {F{\"u}l{\"o}p}, {Gervino}, {Guglielmetti}, {Gustavino},
  {Gy{\"u}rky}, {Jesus}, {Junker}, {Lemut}, {Menegazzo}, {Prati}, {Roca},
  {Rolfs}, {Romano}, {Rossi Alvarez}, {Sch{\"u}mann}, {Somorjai}, {Straniero},
  {Strieder}, {Terrasi}, {Trautvetter}, {Vomiero}, \&
  {Zavatarelli}}]{Formicola2004}
{Formicola}, A., {Imbriani}, G., {Costantini}, H., {et~al.} 2004, Physics
  Letters B, 591, 61

\bibitem[{{Fossat} {et~al.}(2017){Fossat}, {Boumier}, {Corbard}, {Provost},
  {Salabert}, {Schmider}, {Gabriel}, {Grec}, {Renaud}, {Robillot},
  {Roca-Cort{\'e}s}, {Turck-Chi{\`e}ze}, {Ulrich}, \& {Lazrek}}]{Fossat2017}
{Fossat}, E., {Boumier}, P., {Corbard}, T., {et~al.} 2017, \aap, 604, A40

\bibitem[{{Fuller} {et~al.}(2014){Fuller}, {Lecoanet}, {Cantiello}, \&
  {Brown}}]{Fuller2014}
{Fuller}, J., {Lecoanet}, D., {Cantiello}, M., \& {Brown}, B. 2014, \apj, 796,
  17

\bibitem[{{Gallet} \& {Bouvier}(2013)}]{Gallet2013}
{Gallet}, F. \& {Bouvier}, J. 2013, \aap, 556, A36

\bibitem[{{Garc{\'{\i}}a} {et~al.}(2007){Garc{\'{\i}}a}, {Turck-Chi{\`e}ze},
  {Jim{\'e}nez-Reyes}, {Ballot}, {Pall{\'e}}, {Eff-Darwich}, {Mathur}, \&
  {Provost}}]{Garcia2007}
{Garc{\'{\i}}a}, R.~A., {Turck-Chi{\`e}ze}, S., {Jim{\'e}nez-Reyes}, S.~J.,
  {et~al.} 2007, Science, 316, 1591

\bibitem[Gehan et al.(2016)]{Gehan2016} Gehan, C., Mosser, B., \& Michel, E.\ 2016, arXiv:1612.05414

\bibitem[{{Iglesias} \& {Rogers}(1996)}]{Iglesias1996}
{Iglesias}, C.~A. \& {Rogers}, F.~J. 1996, \apj, 464, 943

\bibitem[{{Irwin} {et~al.}(2008){Irwin}, {Hodgkin}, {Aigrain}, {Bouvier},
  {Hebb}, {Irwin}, \& {Moraux}}]{Irwin2008a}
{Irwin}, J., {Hodgkin}, S., {Aigrain}, S., {et~al.} 2008, \mnras, 384, 675

\bibitem[{{Keen} {et~al.}(2015){Keen}, {Bedding}, {Murphy}, {Schmid}, {Aerts},
  {Tkachenko}, {Ouazzani}, \& {Kurtz}}]{Keen2015}
{Keen}, M.~A., {Bedding}, T.~R., {Murphy}, S.~J., {et~al.} 2015, \mnras, 454,
  1792

\bibitem[{{Kurtz} {et~al.}(2014){Kurtz}, {Saio}, {Takata}, {Shibahashi},
  {Murphy}, \& {Sekii}}]{Kurtz2014}
{Kurtz}, D.~W., {Saio}, H., {Takata}, M., {et~al.} 2014, \mnras, 444, 102

\bibitem[{{Ledoux}(1951)}]{Ledoux1951}
{Ledoux}, P. 1951, \apj, 114, 373

\bibitem[{{Maeder} \& {Zahn}(1998)}]{Maeder1998}
{Maeder}, A. \& {Zahn}, J. 1998, \aap, 334, 1000

\bibitem[{{Marques} {et~al.}(2013){Marques}, {Goupil}, {Lebreton}, {Talon},
  {Palacios}, {Belkacem}, {Ouazzani}, {Mosser}, {Moya}, {Morel}, {Pichon},
  {Mathis}, {Zahn}, {Turck-Chi{\`e}ze}, \& {Nghiem}}]{Marques2013}
{Marques}, J.~P., {Goupil}, M.~J., {Lebreton}, Y., {et~al.} 2013, \aap, 549,
  A74

\bibitem[{{Mathis} {et~al.}(2004){Mathis}, {Palacios}, \& {Zahn}}]{Mathis2004b}
{Mathis}, S., {Palacios}, A., \& {Zahn}, J.-P. 2004, \aap, 425, 243

\bibitem[{{Mathis} \& {Zahn}(2004)}]{Mathis2004}
{Mathis}, S. \& {Zahn}, J. 2004, \aap, 425, 229

\bibitem[{{Matt} {et~al.}(2015){Matt}, {Brun}, {Baraffe}, {Bouvier}, \&
  {Chabrier}}]{Matt2015}
{Matt}, S.~P., {Brun}, A.~S., {Baraffe}, I., {Bouvier}, J., \& {Chabrier}, G.
  2015, \apjl, 799, L23

\bibitem[{{Matt} {et~al.}(2012){Matt}, {MacGregor}, {Pinsonneault}, \&
  {Greene}}]{Matt2012}
{Matt}, S.~P., {MacGregor}, K.~B., {Pinsonneault}, M.~H., \& {Greene}, T.~P.
  2012, \apjl, 754, L26

\bibitem[{{Moraux} {et~al.}(2013){Moraux}, {Artemenko}, {Bouvier}, {Irwin},
  {Ibrahimov}, {Magakian}, {Grankin}, {Nikogossian}, {Cardoso}, {Hodgkin},
  {Aigrain}, \& {Movsessian}}]{Moraux2013}
{Moraux}, E., {Artemenko}, S., {Bouvier}, J., {et~al.} 2013, \aap, 560, A13

\bibitem[{{Morel}(1997)}]{Morel1997}
{Morel}, P. 1997, \aaps, 124, 597

\bibitem[{{Morel} \& {Lebreton}(2008)}]{Morel2008}
{Morel}, P. \& {Lebreton}, Y. 2008, \apss, 316, 61

\bibitem[{{Mosser} {et~al.}(2012){Mosser}, {Goupil}, {Belkacem}, {Marques},
  {Beck}, {Bloemen}, {De Ridder}, {Barban}, {Deheuvels}, {Elsworth}, {Hekker},
  {Kallinger}, {Ouazzani}, {Pinsonneault}, {Samadi}, {Stello}, {Garc{\'{\i}}a},
  {Klaus}, {Li}, {Mathur}, \& {Morris}}]{Mosser2012b}
{Mosser}, B., {Goupil}, M.~J., {Belkacem}, K., {et~al.} 2012, \aap, 548, A10

\bibitem[{{Murphy} {et~al.}(2016){Murphy}, {Fossati}, {Bedding}, {Saio},
  {Kurtz}, {Grassitelli}, \& {Wang}}]{Murphy2016}
{Murphy}, S.~J., {Fossati}, L., {Bedding}, T.~R., {et~al.} 2016, \mnras, 459,
  1201

\bibitem[{{Ouazzani} {et~al.}(2012){Ouazzani}, {Dupret}, \&
  {Reese}}]{Ouazzani2012b}
{Ouazzani}, R.-M., {Dupret}, M.-A., \& {Reese}, D.~R. 2012, \aap, 547, A75

\bibitem[{{Ouazzani} {et~al.}(2015){Ouazzani}, {Roxburgh}, \&
  {Dupret}}]{Ouazzani2015}
{Ouazzani}, R.-M., {Roxburgh}, I.~W., \& {Dupret}, M.-A. 2015, \aap, 579, A116

\bibitem[{{Ouazzani} {et~al.}(2017){Ouazzani}, {Salmon}, {Antoci}, {Bedding},
  {Murphy}, \& {Roxburgh}}]{Ouazzani2017}
{Ouazzani}, R.-M., {Salmon}, S.~J.~A.~J., {Antoci}, V., {et~al.} 2017, \mnras,
  465, 2294

\bibitem[{{Peimbert} {et~al.}(2007){Peimbert}, {Luridiana}, \&
  {Peimbert}}]{Peimbert2007}
{Peimbert}, M., {Luridiana}, V., \& {Peimbert}, A. 2007, \apj, 666, 636

\bibitem[{{Rogers} {et~al.}(1996){Rogers}, {Swenson}, \&
  {Iglesias}}]{Rogers1996}
{Rogers}, F.~J., {Swenson}, F.~J., \& {Iglesias}, C.~A. 1996, \apj, 456, 902

\bibitem[{{Royer}(2009)}]{Royer2009}
{Royer}, F. 2009, in Lecture Notes in Physics, Berlin Springer Verlag, Vol.
  765, The Rotation of Sun and Stars, 207--230

\bibitem[{{Saio} {et~al.}(2015){Saio}, {Kurtz}, {Takata}, {Shibahashi},
  {Murphy}, {Sekii}, \& {Bedding}}]{Saio2015}
{Saio}, H., {Kurtz}, D.~W., {Takata}, M., {et~al.} 2015, \mnras, 447, 3264

\bibitem[{{Schatzman}(1962)}]{Schatzman1962}
{Schatzman}, E. 1962, Annales d'Astrophysique, 25, 18

\bibitem[{{Schmid} {et~al.}(2015){Schmid}, {Tkachenko}, {Aerts}, {Degroote},
  {Bloemen}, {Murphy}, {Van Reeth}, {P{\'a}pics}, {Bedding}, {Keen}, {Pr{\v
  s}a}, {Menu}, {Debosscher}, {Hrudkov{\'a}}, {De Smedt}, {Lombaert}, \&
  {N{\'e}meth}}]{Schmid2015}
{Schmid}, V.~S., {Tkachenko}, A., {Aerts}, C., {et~al.} 2015, \aap, 584, A35

\bibitem[{{Schou} {et~al.}(1998){Schou}, {Antia}, {Basu}, {Bogart}, {Bush},
  {Chitre}, {Christensen-Dalsgaard}, {di Mauro}, {Dziembowski}, {Eff-Darwich},
  {Gough}, {Haber}, {Hoeksema}, {Howe}, {Korzennik}, {Kosovichev}, {Larsen},
  {Pijpers}, {Scherrer}, {Sekii}, {Tarbell}, {Title}, {Thompson}, \&
  {Toomre}}]{Schou1998}
{Schou}, J., {Antia}, H.~M., {Basu}, S., {et~al.} 1998, \apj, 505, 390

\bibitem[{{Talon} \& {Charbonnel}(2005)}]{Talon2005}
{Talon}, S. \& {Charbonnel}, C. 2005, \aap, 440, 981

\bibitem[{{Talon} {et~al.}(1997){Talon}, {Zahn}, {Maeder}, \&
  {Meynet}}]{Talon1997}
{Talon}, S., {Zahn}, J., {Maeder}, A., \& {Meynet}, G. 1997, \aap, 322, 209

\bibitem[{{Tassoul}(1980)}]{Tassoul1980}
{Tassoul}, M. 1980, \apjs, 43, 469

\bibitem[{{Unno} {et~al.}(1989){Unno}, {Osaki}, {Ando}, {Saio}, \&
  {Shibahashi}}]{Unno1989}
{Unno}, W., {Osaki}, Y., {Ando}, H., {Saio}, H., \& {Shibahashi}, H. 1989,
  {Nonradial oscillations of stars}

\bibitem[{{Van Reeth} {et~al.}(2016){Van Reeth}, {Tkachenko}, \&
  {Aerts}}]{VanReeth2016}
{Van Reeth}, T., {Tkachenko}, A., \& {Aerts}, C. 2016, \aap, 593, A120

\bibitem[{{Van Reeth} {et~al.}(2015){Van Reeth}, {Tkachenko}, {Aerts},
  {P{\'a}pics}, {Triana}, {Zwintz}, {Degroote}, {Debosscher}, {Bloemen},
  {Schmid}, {De Smedt}, {Fremat}, {Fuentes}, {Homan}, {Hrudkova},
  {Karjalainen}, {Lombaert}, {Nemeth}, {{\O}stensen}, {Van De Steene}, {Vos},
  {Raskin}, \& {Van Winckel}}]{VanReeth2015}
{Van Reeth}, T., {Tkachenko}, A., {Aerts}, C., {et~al.} 2015, \apjs, 218, 27

\bibitem[{{Venuti} {et~al.}(2017){Venuti}, {Bouvier}, {Cody}, {Stauffer},
  {Micela}, {Rebull}, {Alencar}, {Sousa}, {Hillenbrand}, \&
  {Flaccomio}}]{Venuti2017}
{Venuti}, L., {Bouvier}, J., {Cody}, A.~M., {et~al.} 2017, \aap, 599, A23

\bibitem[{{Zahn}(1992)}]{Zahn1992}
{Zahn}, J. 1992, \aap, 265, 115

\end{thebibliography}

\end{document}